\PassOptionsToPackage{dvipsnames}{xcolor}
\documentclass{aa}

\usepackage{graphicx}
\graphicspath{{figures/}}
\usepackage[varg]{txfonts}
\usepackage{commath}
\usepackage[super]{nth}
\usepackage{natbib}
\usepackage[most]{tcolorbox}
\usepackage{soulpos}
\usepackage{xcolor}
\bibpunct{(}{)}{;}{a}{}{,}

\usepackage{siunitx}
\usepackage[dvipsnames]{xcolor}
\definecolor{click_color}{RGB}{46,48,118}
\usepackage{hyperref}
\hypersetup{
  colorlinks=true,
  linkcolor=click_color,
  citecolor=click_color,
  urlcolor=click_color
  }
\usepackage[capitalize]{cleveref}
\crefname{equation}{eq.}{eqs.}
\crefname{section}{Sect.}{Sects.}

\begin{document} 
   \title{Observing radio transients with Phased ALMA: Pulses from the Galactic Centre magnetar}
   \author{J. Vera-Casanova\thanks{Email: josefina.vera@alumni.uc.cl}
          \inst{1,2}
          \and M. Cruces \inst{1,2,3,4,5}
          \and K. Liu \inst{5,6}
          \and J. Wongphechauxsorn \inst{7,5}
          \and C.A. Braga \inst{1,8}
          \and M. Kramer \inst{5}
          \and P. Torne \inst{9,5}
          \and P. Limaye  \inst{5,10}
          \and M.C. Espinoza-Dupouy \inst{11}
          \and L. Rodriguez \inst{8}
          }

\institute{
    Centre of Astro-Engineering, Pontificia Universidad Católica de Chile, Av. Vicuña Mackenna 4860, Santiago, Chile
    \and
    Department of Electrical Engineering, Pontificia Universidad Católica de Chile, Av. Vicuña Mackenna 4860, Santiago, Chile
    \and
    Joint ALMA Observatory, Alonso de Córdova 3107, Vitacura, Santiago, Chile
    \and
    European Southern Observatory, Alonso de Córdova 3107, Vitacura, Casilla 19001, Santiago de Chile, Chile
    \and
    Max-Planck-Institut für Radioastronomie, Auf dem Hügel 69, D-53121 Bonn, Germany
     \and
    Shanghai Astronomical Observatory, Chinese Academy of Sciences. Xujiahui, Xuhui District, Shanghai, China
    \and
     Julius-Maximilians-Universität Würzburg, Institut für Theoretische Physik und Astrophysik, Lehrstuhl für Astronomie, Emil-Fischer-Straße 31, D-97074 Würzburg, Germany
     \and
     Instituto de Astrofísica, Facultad de Física, Pontificia Universidad Católica de Chile, Casilla 306, Santiago 22, Chile
    \and 
    Institut de Radioastronomie Millimétrique, Avda. Divina Pastora 7, Local 20, E-18012 Granada, Spain
    \and
    Argelander Institute for Astronomy, 53121 Bonn, Germany
    \and
    Departament of Astronomy, Universidad de Chile, Camino El Observatorio 1515, Las Condes, Santiago, Chile
    }
   \date{Received 08 April 2025; accepted 15 July 2025}
 
\abstract
{Radio transients, such as pulsars and fast radio bursts (FRBs), are primarily detected at centimetre (cm) radio wavelengths, where the highest luminosities are found. However, observations of sources in dense environments are heavily affected by propagation effects, such as scattering, which may hinder a detection. Millimetre (mm)-wave observations bypass this complication but require the largest radio telescopes to compensate for the lower flux densities. When used in phased mode, the ALMA radio telescope provides an equivalent dish size of $\sim$84m, making it the most sensitive instrument at mm/sub-mm wavelengths. In combination with its high time resolution, it offers a unique opportunity to study radio transients in an unexplored frequency window.
}
    {We studied the Galactic Centre (GC) magnetar, PSR J1745-2900, as a laboratory for magnetars in complex magneto-turbulent environments and for linking with FRBs. Through this pilot study, we showcase the potential of ALMA in its phased configuration to observe radio transients and to achieve, for some sources, the first ever detections outside the cm-wave range.
   }
   {We studied the GC magnetar using ALMA archival data of Sgr A* at Band 3, taken during the 2017 GMVA campaign. The data were searched in intensity, and the pulses were classified based on their circular and linear polarisation properties and arrival phase.
   }
   {We detected eight highly polarised pulses from the GC magnetar with energies in the range of $10^{29}$ erg. We constructed its cumulative energy distribution and we fit a power law, assuming the event rate scales with the energy as $R \propto E^{\gamma}$. The result is an exponent of $\gamma = -2.4 \pm 0.1$, which is consistent with values reported for magnetars at cm-waves and repeating FRBs. With the $\gamma$-value and the system properties of the phased ALMA mode, we estimate that over 160 known pulsars could be detected by ALMA. For repeating FRBs, observing during their peak activity window could lead to detections of several bursts per hour.
   }
  {We expect that ALMA's lower frequency bands with polarisation capabilities, will serve as a pioneer on mm–wave searches for pulsars and to study complex environments involving radio transients.
  }
   \maketitle
\section{Introduction}

A particular type of radio emitting neutron stars (NSs) are magnetars. They are young NSs ($\sim10^{3}$ yr ; \citealt{magns}) with extremely strong magnetic fields, of the order of \SIrange{e14}{e15} G \citep{magnetars}. Although more than 30 magnetars have been reported \citep{magnetarcat}, only a handful exhibit detectable emission at radio frequencies. Recently, they have attracted significant attention as promising candidates for the origin of fast radio bursts (FRBs; \citealt{circularpol}, \citealt{magnetoionic}). FRBs were first discovered while inspecting archival data of pulsar surveys \citep{Lorimer}. Since then, over 4000 FRBs have been discovered \citep{secondchime}, the majority classified as one-off events and only 5\% as repeaters \citep{petroff}. These radio transient events, of extragalactic origin, are characterised by their high inferred luminosities ($\rm{\nu L_{\nu} \sim 10^{43}erg\hspace{0.1cm}s^{-1}}$, \citealt{locatelli}) and pulses as short as microseconds \citep{snelders}. Although there is no widely accepted explanation for the origin of FRBs, one of the strongest pieces of evidence that they could be magnetars at cosmological distances is the detection of two bright radio bursts from the Galactic magnetar SGR 1935+2154. These bursts exceeded the typical isotropic energy of regular radio magnetars by approximately three orders of magnitude \citep{Bochenek,frblike}, but are still relatively weak compared to FRBs.

Commonly, the observations of radio transient sources, such as pulsars and FRBs, are performed using large radio telescopes operating at centimetre (cm) wavelengths. This is because these wavelengths yield higher flux densities \citep{machester}. Pulsars, for instance, are known to have a steep spectral index around $\alpha$ = -1.6 \citep{spindex}. However, some magnetars have flatter or inverted values (e.g., \citealt{torne2015, torne2016}). The detectable flux densities are also affected by interstellar effects such as scintillation, scattering and dispersion. The last of these effects introduces a frequency-dependent delay in the arrival of the signal, given by
\begin{equation}
    \Delta t = \rm{k_{DM}} \times \left(\frac{1}{{\nu_{low}}^{2}} - \frac{1}{{\nu_{high}}^{2}} \right) \times \rm{DM}\hspace{0.2cm},
	\label{eq:dispersion}
\end{equation}
where $\rm{k_{DM}} = 1/(2.41 \cdot 10^{-4}) \hspace{0.1cm} pc^{-1}\hspace{0.1cm}cm^{3} MHz^{2}$s, $\rm{\nu_{low}}$ and $\rm{\nu_{high}}$ are the lower and higher frequencies of the observing band, respectively \citep{handbook}. DM refers to the dispersion measure, and quantifies the integrated column density of electrons along the line of sight
\begin{equation}
    \rm{DM} = \int_{0}^{d}\rm{n_{e}}dl\hspace{0.2cm},
\end{equation}
where $\rm{n_e}$ is the electron number density and $\rm{d}$ is the column length (distance). 
 
 From \href{eq:dispersion}{Eq. 1}, by fixing the DM, it is evident that the lower the frequency, the higher the dispersion delay. In particular, for a source with $\rm{DM}$ = 1770 $\rm{pc\hspace{0.1cm}cm^{-3}}$, observing at 1.4 GHz the delay is approximately 1.793 s, while at 42 GHz it is 366 $\mu$s, and at 86 GHz it is 46 $\mu$s (see \href{tab:delay_table}{Table 1}). 

\begin{table*}
\begin{center}

\caption{Dispersion delay ($\rm{\Delta t}$) for different observing set-up configurations.}
\label{tab:delay_table}
\begin{tabular}{ c  c  c  c  c  c  c  c }
\hline
Instrument  & $\rm{\nu_{low}}$ (GHz) & $\rm{\nu_{high}}$ (GHz) & $\rm{\nu_{c}}$ (MHz) & $\rm{\Delta \nu}$ (MHz) & $\rm{t_{samp}}$ ($\mu$s) &$\rm{\Delta t}$ ($\mu$s) & samples delay \\ \hline
Effelsberg (L-band) & 1.210 & 1.510 & 1.360 & 300 & 54.6 & 1793000 & 32845  \\
ALMA (Band 1)  & 42.168 & 44.168 & 43.168 & 2000 & 32 & 366 & 11  \\
ALMA (Band 3)  & 85.268 & 87.268 & 86.268 & 2000 & 32 & 46 & 1    \\
\hline
\end{tabular}
\tablefoot{We assume a DM = 1770 $\rm{pc\hspace{0.1cm}cm^{-3}}$. $\rm{\nu_{low}}$ and $\rm{\nu_{high}}$ are the lower and higher frequency of the bandwidth ($\rm{\Delta \nu}$). The central frequency of the receiver is $\rm{\nu_{c}}$ and $\rm{t_{samp}}$ is the sample time. For ALMA Band 1, we assume the same sample time as Band 3. At the final column is given the delay in number of samples.}
\end{center}
\label{tab:tab1}
\end{table*}

Due to its complex and scattered medium, the vicinity of Sgr A$^{*}$ holds the potential to be used as a laboratory for testing the behaviour of pulsars and magnetars in extreme environments and the ability to link them with FRBs.  The DM excess and highly varying rotation measure (RM) of FRBs indicate that they could be born in complex magneto-turbulent environments \citep{platts2019living,mottez2020,wang2022}. Similar properties could be found around pulsars and the magnetar near the Galactic Centre (GC, \citealt{desvignesgc}, \citealt{abbate}). For instance, large RMs are expected in the boundaries of massive black holes \citep{bowersgra,magnetoionic}. Importantly, given the high plasma densities of regions such as the GC ($\rm{n_{e} \approx 4 \cdot 10^{7} cm^{-3}}$;  \citealt{witzel}), propagation effects such as the dispersion delay and scattering may hinder a detection at low radio frequencies \citep{torne2023}. For scattering, the timescale is inversely proportional to the fourth power of the frequency as \rm{$\tau_{s} \approx \nu^{-4}$} \citep{handbook}. One way to overcome these two propagation effects and reach the densest regions is to observe at higher frequencies. To compensate for the lower flux densities, the largest possible radio telescopes are required.
 
The ALMA Phasing System (APS, \citealt{matthews2017}) was developed for very long baseline interferometry (VLBI) observations in global campaigns such as the Global Millimetre VLBI Array (GMVA, \citealt{data}) and the Event Horizon Telescope (EHT, \citealt{eht}). APS is available with an odd number of antennas as a hardware requirement \citep{matthews2017}. First it was implemented in Band 6 and Band 3 \citep{goddialmadata} and then at Band 7 \citep{phasing2023} and Band 1. When combining 49 of its 12m antennas, its equivalent dish size is $\approx$ 84m, which makes it the most sensitive instrument at millimetre (mm) to sub-millimetre (sub-mm) wavelengths. Previous observations of known pulsars with phased ALMA have detected sources such as the Vela Pulsar \citep{Liu_2019}. Searches for new pulsars have also been conducted in recent years \citep{Liu_2021,torne2023}. For FRBs, no observations have been performed with phased ALMA to date. 

Based on APS, the Phased ALMA mode (PAM) was introduced in Cycle 8 (2021) to enable observations of weak radio sources (<50 mJy) in time domain. The phasing of ALMA antennas involves multiple steps \citep{goddialmadata}. First, phase corrections align each signal to a reference antenna. Then, the signals of each antenna are coherently summed and the raw voltage data are recorded as 2 bit samples using MARK 6 recorders. When the output is used for pulsar searching, the data are packeted and divided in frequency channels, with a time resolution multiple of 8 $\mu$s \citep{almahand}. 

For this work we studied the GC magnetar, PSR J1745-2900, using ALMA archival data of Sgr A$^{*}$. This is because the source is located within the same synthesised beam of Sgr A$^{*}$ ($\theta \approx$ 5.7''), due to its 0.07-2 pc projected distance \citep{reapsr}. We searched for the magnetar --single pulses to link them to FRB emission. We examined the potential of phased ALMA to open a window to higher frequency observations of radio transients. This paper is organized as follows. In Section \ref{sec:sec2} we describe the observations, and then in Section \ref{sec:sec3} the single pulse search. In Section \ref{sec:sec4} we present the results of the pulses characterisation. In Section \ref{sec:sec5} we discuss the future prospects of using phased ALMA for radio transient observations. Finally, in Section \ref{sec:sec6} we present our final remarks.

\section{Observations}
\label{sec:sec2}

We used observations of Sgr A* from the 2017 GMVA campaign at 3.5 mm (Project Code: 2017.1.00795.V, PI: M. Johnson). The observations were carried out on April 3, using 37 12m antennas in Band 3 configuration centred at 86 GHz \citep{Liu_2021}. The dataset covered a total of 5.2 hr, with on-source time of 2.5 hr. We used a total of $\sim$2 hr observing time, excluding 30 minutes of unusable data (see \href{tab:table1}{Table 2}). The off-source time corresponds to the calibrations needed between source scans. Some calibrators were 3C 279, NRAO 530 and J1924-2914 (OV-236).

As discussed in \cite{Liu_2021}, the data was converted to PSRFITS format \citep{psrfits} with full Stokes information (I,Q,U,V) and 32 frequency channels, each with a 62.5 MHz channel width. The outputs were multiple files corresponding to each scan, with an approximate duration of 52 s. Every 18.192 s there was a drop-off on the time series caused by the phasing cycle of the array. This artefact was previously reported in \cite{Liu_2021}. We list the observations in \href{tab:obs}{Table 2}.

\begin{table}
\begin{center}
\caption{Observations of Sgr A$^{*}$ used in this work, taken during the 2017 GMVA campaign.}
\label{tab:table1}
\begin{tabular}{ c c c }
\hline
Scan   &  Start time (UTC) & End time (UTC) \\ \hline
No0133 & 2017-04-03 07:46:05 & 2017-04-03 07:53:05 \\
No0134 & 2017-04-03 07:55:05 & 2017-04-03 08:02:05 \\
No0142 & 2017-04-03 08:16:13 & 2017-04-03 08:23:13 \\
No0143 & 2017-04-03 08:25:13 & 2017-04-03 08:32:13 \\
No0152 & 2017-04-03 08:46:22 & 2017-04-03 08:53:22 \\
No0153 & 2017-04-03 08:55:22 & 2017-04-03 09:02:22 \\
No0217 & 2017-04-03 12:17:08 & 2017-04-03 12:24:08 \\
No0220 & 2017-04-03 12:26:08 & 2017-04-03 12:33:08 \\
No0227 & 2017-04-03 12:47:05 & 2017-04-03 12:54:05 \\
No0230 & 2017-04-03 12:56:05 & 2017-04-03 13:03:05 \\
No0237 & 2017-04-03 13:17:01 & 2017-04-03 13:24:01 \\
No0240 & 2017-04-03 13:26:01 & 2017-04-03 13:33:01\\
No0242 & 2017-04-03 13:38:31 & 2017-04-03 13:45:31\\
No0243 & 2017-04-03 13:47:01 & 2017-04-03 13:54:01\\
No0250 & 2017-04-03 14:12:01 & 2017-04-03 14:19:01\\
No0252 & 2017-04-03 14:24:31 & 2017-04-03 14:31:31\\
No0253 & 2017-04-03 14:33:01 & 2017-04-03 14:40:01\\
\hline
\end{tabular}
\tablefoot{The data is centered at 86.268 GHz with 2 GHz of bandwidth. Each scan consists of multiple PSRFITS files containing full Stokes information.}
\end{center}
\end{table}

\section{Methods and data processing}
\label{sec:sec3}

We searched for single pulses using a PRESTO-based pipeline  \citep{presto}, detailed in \cite{Braga2024} with some modifications to account for PSRFITS containing Stokes. We used a signal-to-noise ratio (S/N) threshold of 6, determined through analysis of the noise floor of the time series on pulse-free regions. These time series were constructed by summing all the frequency channels of the PSRFITS files on Stokes I. Given the high DM of the source and the high observing frequency at Band 3, the dispersion delay between the highest and the lowest frequency channel is 46 $\mu$s (see \href{tab:delay_table}{Table 1}). Therefore, DM trials were unnecessary, and we adopted the reported value of the source, 1770 $\rm{pc \hspace{0.1cm}cm^{-3}}$ \citep{Liu_2021}. Using this value we searched for single pulses in Stokes I. We inspected each candidate visually. To this end, we constructed the time series in intensity, linear and circular polarisation.

Importantly, due to the negligible dispersion delay, we relied primarily on polarisation to confirm the pulse candidates. We complemented these data with a second test --in which we used the spin period of the magnetar to assign a phase to the arrival time of each pulse. This arrival phase was compared with the region where the pulse is expected to arrive. This region was determined from the full width of the integrated pulse profile. We used this method as most single pulses are significantly narrower than their integrated pulse, which is built up from adding up thousands of pulses \citep{drakepulses}. Nevertheless, in some cases magnetars also show single pulses outside the main profile \citep{camiloxte}.

According to our criteria, if a pulse candidate has significant linear and circularly polarised flux and it arrives in phase with the integrated pulse profile of PSR J1745$-$2900, then it is considered a real pulse. It is worth noting that pulses with small polarisation fractions would have not been detected by our pipelines. 

\section{Results}\label{sec:sec4}

We report eight single pulses from the GC magnetar, PSR J1745-2900, which are shown in \href{fig:pulses}{Fig. 1}. The bottom panel shows the dynamic spectra across all 32 frequency channels. The top panel displays the integration along the bandwidth, we also refer to this as the single pulse profile. We performed this integration using Stokes I. The pulses, together with their linearly and circularly polarized flux, are shown with respect to their arrival phase in \href{fig:pulses_phase}{Fig. 2}. We calculated the time of arrival (TOA) of each reported pulse. In our search we found the three pulses reported by \cite{Liu_2021} plus five new pulses.

Using all data blocks from \href{tab:table1}{Table 2}, we created integrated pulse profiles per scan. From those with the highest S/N, we obtained TOAs that we input to TEMPO \citep{tempo}, following standard timing procedures. This step was necessary as the ephemeris listed in the ATNF catalogue \citep{catalogue} was not sufficient to fully recover the signal of the magnetar. Through residual minimisation, with the spin period as the only free parameter, we generated a new ephemeris file shown in \href{tab:timing4}{Table 3}. 

\begin{table}
\begin{center}
\caption{Ephemeris File.}\label{tab:timing4}
\begin{tabular}{ll}
\hline
& PSR J1745$-$2900    \\
\hline
\hline
Right ascension, $\alpha$ (J2000)\dotfill & 17:45:40.1662(8) \\
Declination, $\delta$ (J2000)\dotfill & $-$29:00:29.89(1)  \\
Spin frequency, $F$ ($\text{s}^{-1}$)\dotfill & 0.26762(8)\\
Dispersion measure, DM ($\text{cm}^{-3}\,\text{pc}$)\dotfill &  1770(3)\\
Reference Epoch, PEPOCH \dotfill &  57846.349663 \\
\hline
\hline
\end{tabular}
\tablefoot{The DM uncertainty was taken from the ATNF catalogue \citep{catalogue}.}
\end{center}
\end{table}

\begin{figure*}
    \centering
	\includegraphics[width=18.5cm]{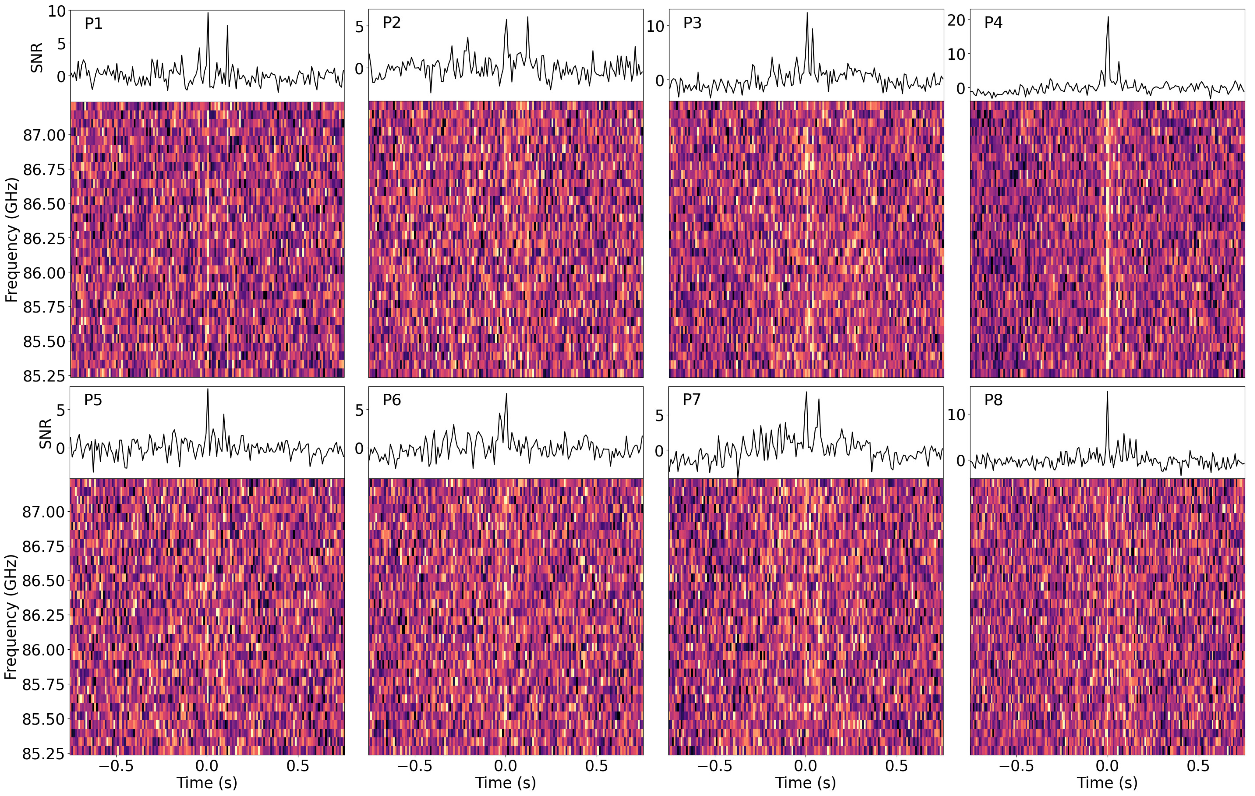}
    \caption{Pulses of the GC magnetar, PSR J1745-2900, found in the Sgr A* GMVA 2017 campaign. For each plot, the top panel shows the pulse profile in intensity, and the bottom panel displays the dynamic spectra along APS bandwidth at Band 3. The spectral resolution is given by the 32 frequency channels across the 2 GHz bandwidth and the time resolution was set to 9.6 ms for visualisation purposes.}
    \label{fig:new_pulses}
\end{figure*}

\begin{figure}
    \centering
	\includegraphics[width=9cm]{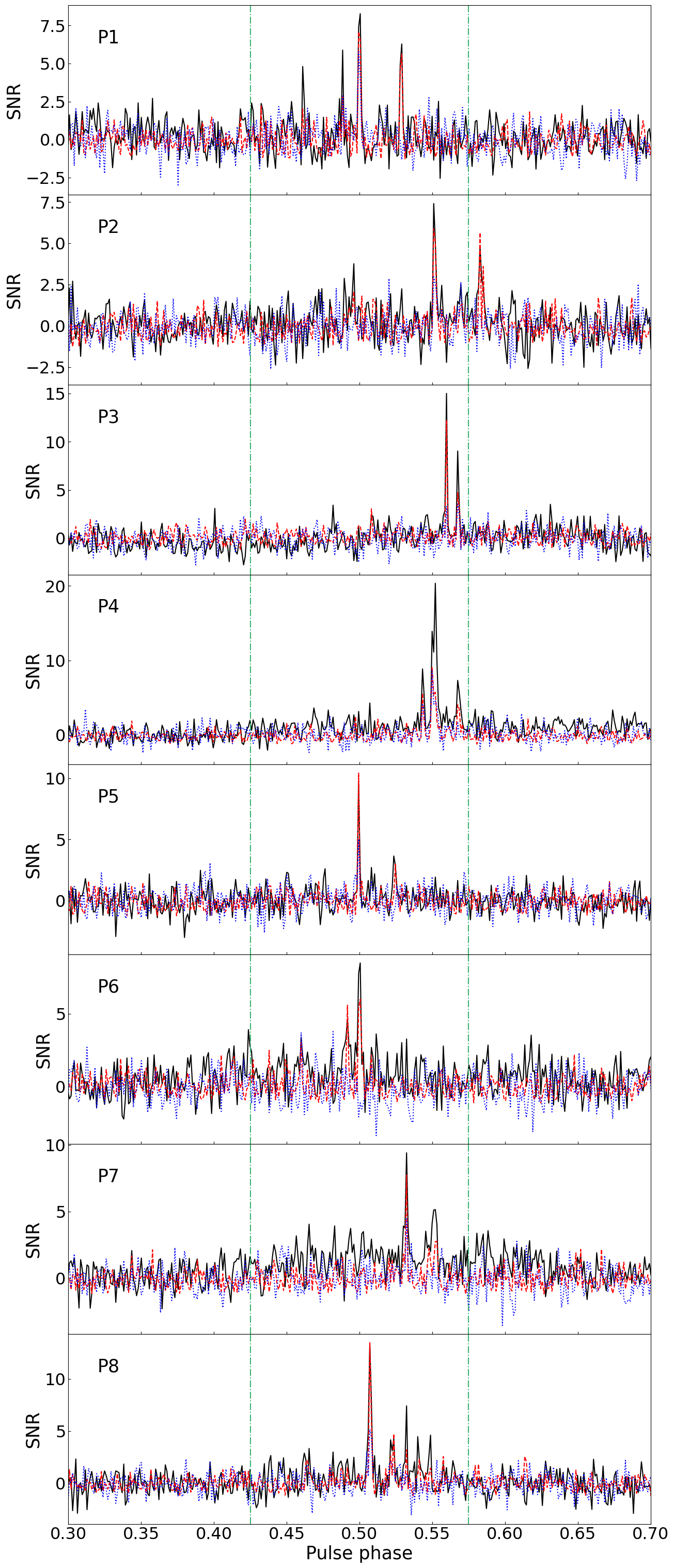}
    \caption{Arrival phase of the pulses from PSR J1745-2900. Each pulse profile is displayed in intensity (Stokes I), linear (L) and circular (V) polarisation in black, red and blue, respectively. The time series has been downsampled by a factor of 128 (4.1 ms of time resolution). The green dashed vertical lines show the regions where the pulses are expected to arrive. We calculated the TOAs of each reported pulse, with respect to the barycentre and infinite frequency.}
    \label{fig:pulses_phase}
\end{figure}

\subsection{Pulse characterisation}

In \href{tab:pulses_table}{Table 4} we report the pulse properties. We report the modified Julian day (MJD), S/N, full width at half maximum (FWHM), flux density, fluence \footnote{Corresponding to the band-averaged peak flux density.} and isotropic energy. As shown in \href{fig:new_pulses}{Fig. 1}, most of the pulses from the GC magnetar have more than one component. We computed the S/N by normalising the time series using a pulse-free region and taking the pulse peak. To estimate the pulse width, we used a multi-Gaussian fitting routine developed for single pulse analysis (Limaye et al. in prep.). Then, the fluence $\rm{F}$ is obtained as
\begin{equation}
    F = \frac{\mathrm{S/N} \times \mathrm{SEFD}}{\sqrt{n_{\mathrm{pol}} \times n_{\mathrm{sp}} \times \Delta \nu}} \sqrt{\mathrm{FWHM}} \hspace{0.2cm},
    \label{eq:radiometer}
\end{equation}
where $\rm{n_{pol}}$ is the number of polarisations, $\rm{n_{sp}}$ the number of spectral windows and $\Delta \nu$ the bandwidth of each spectral window. The system equivalent flux density (SEFD) is a parameter given for each telescope as
\begin{equation}
    \rm{SEFD = \frac{2 k_{b} \times T_{sys}}{A_{eff}}= \frac{T_{sys}}{G}}\hspace{0.2cm},
\end{equation}
where $\rm{k_b}$ is the Boltzmann constant, $\rm{T_{sys}}$ is the system temperature, G is the system gain of phased ALMA (51 K and 1.15 K/Jy respectively on \citealt{Liu_2021}), and $\rm{A_{eff}}$ is the effective area. Then, the isotropic energy is given by
\begin{equation}
    \rm{E = F \times \Delta \nu \times 4 \pi D^{2} \times 10^{-23}}\hspace{0.2cm} (erg),
    \label{eq:energy}
\end{equation}
where D corresponds to the source luminosity distance, equal to 8.3 kpc \citep{yim}. As listed in \href{tab:pulses_table}{Table 4}, the energy of the pulses is of the order of $10^{29}$ erg. 

\begin{table*}
	\centering
	\label{tab:pulses_table}
    \caption{Parameters of the pulses from PSR J1745-2900 found in APS at Band 3.}
    \resizebox{14cm}{!}{%
	\begin{tabular}{ c c c c c c c } 
		\hline
		  Candidate & S/N & MJD  (bary) &  FWHM (s) & Flux density (Jy) & Fluence (Jy ms) & Energy (10$^{29}$ erg) \\
		\hline
		1 & 11.0 & 57846.3495 & 0.11 & 0.023 & 2.589 & 4.27 \\
		2 & 6.6  & 57846.5413 & 0.13 & 0.013 & 1.660 & 2.74 \\
		3 & 15.6 & 57846.5730 & 0.04 & 0.058  & 2.068 & 3.41 \\
        4 & 22.2 & 57846.5422 & 0.04 & 0.074  & 3.284 & 5.41 \\
		5 & 7.5  & 57846.3478 & 0.08 & 0.018 & 1.499 & 2.47 \\
		6 & 10.7  & 57846.3493 & 0.04 & 0.036 & 1.556  & 2.57 \\
        7 & 8.9  & 57846.5412 & 0.10 & 0.020  & 1.956 & 3.23 \\
        8 & 14.7 & 57846.3767 &  0.01 & 0.127  & 0.837 & 1.38  \\
		\hline
	\end{tabular}%
    }
    \tablefoot{The S/N was obtained from the pulse profile using a downsampling factor of 128. The FWHM was obtained from a multi-Gaussian fitting, where the width is defined from the S/N peak to the point where the value drops to 50$\%$. Whenever the pulse was fitted with multiple Gaussians, the outermost components were considered for the S/N drop. The flux density and fluence were obtained considering all the 2 GHz bandwidth of the observation.}
\end{table*}

\subsection{Energy distribution}\label{sec:energy_dist}

We constructed the cumulative energy distribution by sorting the events and then counting the number above an energy threshold, until all the pulses were contained. We included only events above the completeness limit \footnote{This is the energy limit of a burst having the minimum detectable fluence given a telescope sensitivity.}, which is $\sim$1120 mJy ms using a $T_{sys} =$ 51 K and $G$ = 1.15 K Jy$^{-1}$ \citep{Liu_2021} and considering the pulse width as the mean FWHM of our pulses, $\approx$ 0.07 s. We fit a power law to the cumulative energy distribution using a least-squares minimisation considering that the event rate scales with the energy as $R \propto E^{\gamma}$. The result is a power-law exponent of $\gamma = -2.4 \pm  0.1$.

\begin{figure}
	\includegraphics[width=\columnwidth]{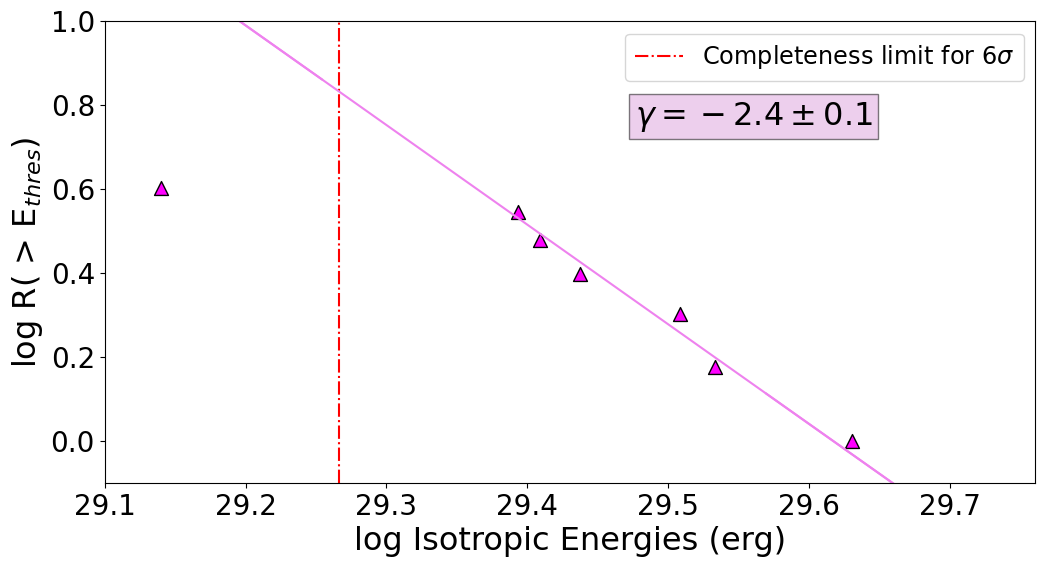}
    \caption{Cumulative energy distribution for the GC magnetar, PSR J1745-2900. The $\gamma$-value was obtained from a least-squared fitting to the events above the completeness limit.}
    \label{fig:energies}
\end{figure}

\section{Discussion}\label{sec:sec5}

Our sample adds five pulses in addition the three previously reported by \cite{Liu_2021}, mainly due to the difference in the candidate selection criteria. In our work, we defined the S/N threshold as 6, based on a statistical analysis of the noise floor. For each of these candidates the four Stokes parameters plus the linear and circular polarisation were inspected. Polarisation information was crucial as the dispersion delay between the frequency channels was negligible at ALMA Band 3 frequencies.

Among the pulses, five of them (P1, P2, P3, P5, and P8) exhibit nearly 100$\%$ of the linearly polarised flux, while the remaining three (P4, P6, and P7) have values close to 50$\%$. In addition, for the majority of them there is a degree of circular polarisation of approximately 30$\%$. Regarding the pulse morphology, as seen in \href{fig:new_pulses}{Fig. 1}, most of the pulses have at least two components, which vary in relative intensity. All pulses show emission across the entire 2 GHz bandwidth. Broadband emission is one of the features of radio pulsars \citep{handbook}.

 This is the first time the energy properties of the pulses from the GC magnetar have been explored at high frequencies. Previous studies were conducted at 3.1 GHz \citep{yan} and 8.6 GHz \citep{yan2015}. Common fits to the energy distribution of single pulses are log-normal and power-law distributions. For the power-law method, which is the one used in our study, the defining parameter is $\gamma$, the power-law exponent such that $R \propto E^\gamma$, with $R$ the rate of events above a given energy value $E$. This definition corresponds to the cumulative energy distribution, which is broadly used in cases where the number of pulses is limited, such as the case of FRBs.
 
As shown in \href{fig:energies}{Fig. 3}, our sample is well fit by $\gamma$ = -2.4 $\pm$ 0.1. For the same source, at cm-wavelengths, the energy distribution of single pulses was found to be well fit by a log-normal distribution with  $\mu$ = 1.34 and $\sigma$ = 0.57 at 3.1\,GHz \citep{yan2015}, and $\mu$ = 0.72 and $\sigma$ = 0.22 at 8.6\,GHz \citep{yan}. The difference with our result was expected given the sample size, where the authors considered 1900+ and 2300+ pulses, respectively. However, it is known that pulsars show a bimodal behaviour: their regular emission is well modelled with a log-normal distribution \citep{Burke-Spolaor2012}, while giant pulses are modelled by a power-law distribution \citep{Ramesh2010}. In the case of the GC magnetar, previous studies on the single pulse peak flux densities of giant pulses at 8.4 GHz, found that they are not well fit by a log-normal distribution \citep{Pearlman_2018}. Given the steep spectral index in pulsars, the observable events at millimetre wavelengths could correspond to the most energetic pulses. Alternatively, a power-law relation for the cumulative energy distribution would be the result of the limited energy range of our pulses. Extending the observations with on-source time and frequency coverage will increase the sample and will provide us with a better general picture of the behaviour of PSR 1745-2900.
 
 Our $\gamma=-2.4\pm0.1$ is consistent with the energy distributions of other magnetar at cm-wavelengths. For XTE J1810$-$197 (PSR J1809$-$1943) giant pulses, values between -2.1 and -7.68 at 1.4 GHz, between -1.78 and -2.86 at 4.9 GHz, and between -0.89 and -1.89 at 8.35 GHz have been reported. The differences could be attributed to its intrinsic magnetar variability and the observations at different frequencies (see Table 2 in \citealt{xte}). 
 
 Importantly, our results fill a critical gap by characterising magnetar radio emission across a scarcely explored frequency window, offering valuable insights into potential changes in their emission mechanisms. Magnetar radio emission--known to be coherent at cm-wavelengths--persists into the mm-band, as demonstrated by our work, by \cite{Liu_2021}, and by \cite{torne2016}. However, it remains unclear exactly where a transition to an incoherent emission occurs. This transition is expected to take place in the mm to sub‑mm range; however, there are currently very few observational facilities equipped for time-domain astronomy at such frequencies. In this context, and given its sensitivity, PAM holds significant potential.

  \begin{figure*}
   \centering
   \includegraphics[width=17cm]{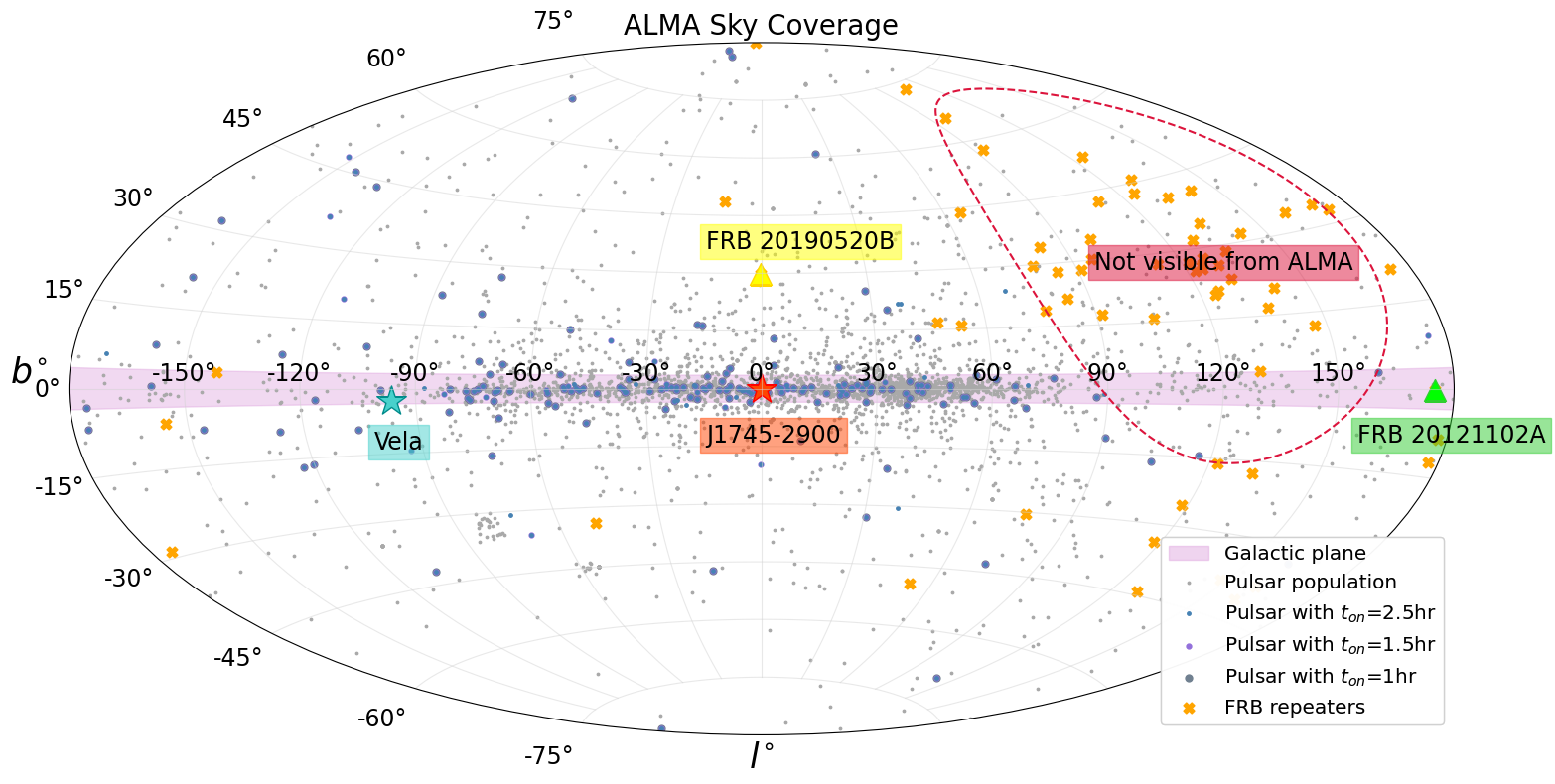}
   \hspace{5cm}\caption{ALMA visibility. The dots correspond to the known pulsar population, where the size of the dots scales inversely with the integration time needed for detection. Larger dots indicate lower integration times. The orange crosses correspond to the repeating FRBs. Some remarkable sources are shown with special icons: stars for pulsars and triangles for FRBs.}
    \label{fig:skyplot}
   \end{figure*}

\subsection{Comparison with FRBs}

In regard to FRBs, there are three sources with seemingly periodic activity windows: FRB~20121102A with a periodicity of 159.3 days \citep{Braga2024}, FRB~20180916B with 16.35 days \citep{chime_periodic}, and the recently reported FRB~20240209A with 126 days \citep{newperiodic}. The periodicity corresponds to the cycle of the  alternating on-and-off activity. Naturally, observations carried out at their peak activity are more likely to result in a detection. FRB 20121102A has $\gamma-$values between $-1.1\pm0.1$ \citep{papermary} and $-1.8\pm0.3$ \citep{Gourdji_2019}. However, \cite{li2} used a large sample of 1600+ bursts detected by the Five-hundred-metre Aperture Spherical Telescope (FAST), and showed that the energy distribution is bimodal and best fit by a combination of a log-normal and a Cauchy function. In that work, the analysis was done over the histogram of the differential energy distribution rather than the cumulative distribution, given the large sample of bursts available. In \cite{li2}, when analysing the high-energy component, it was found that the sub-sample is well fit by a power law with $\gamma=-1.37\pm0.18$.  It is important to note that in our work we reported a limited sample of eight pulses. With a larger sample size (hundreds of pulses) the differential energy distribution could also be fit.

For another active repeater, FRB~20201124A, $\gamma$-values of -1.5 $\pm$ 0.1 (in \citealt{xu}) and from -1.95 $\pm$ 0.01 to -2.25 $\pm$ 0.02 (in \citealt{kristen}) have been reported, both at cm-wavelengths. A steeper value of $\gamma$ is often seen at higher energies in repeating FRBs  \citep{kristen}, reaching values as steep as $\gamma=-4.9$ in the high-energy tail \citep{kumar}. Recent studies of the FRB energy distribution have reported a value of -1.96 $\pm$ 0.15 for the pure power law \citep{frb_energy} using FRB data between 0.83 and 1.6 GHz.

The typical isotropic energies of FRBs are in the range of $10^{36}-10^{41}$ erg \citep{chimemag}. Comparing the energy of our pulses, we find a gap of at least seven orders of magnitude. Nevertheless, the hypothesis of FRBs being extragalactic magnetars \citep{frbarxiv} is supported by the activity of the Galactic magnetar SGR 1935+2154, which emitted bursts with energies in the range of 10$^{34}$ \citep{margalit,Bochenek}. This values are closer to those seen for FRBs, but still at least two orders of magnitude below. There are some phenomena that could explain the energy gap. One is multiple magnifications of the pulses due to plasma lensing, especially in environments where magnetars are orbiting a compact source such as a massive black hole \citep{Pen_2015,Pearlman_2018}. In this regard, a promising laboratory to test this scenario is the GC. While for most radio pulsars and FRBs the monitoring is done at lower frequency bands, complex regions such as the GC are hard to reach with cm-waves as the pulses are heavily affected by propagation effects such as scattering and dispersion. Therefore higher frequencies allow us to probe deeper into these complex regions. The natural disadvantage is the lower luminosities expected at higher frequencies. Our study is especially remarkable given the detection of several single pulses from the GC magnetar at 86 GHz, which allow us to characterise the source, explore the potential of ALMA for radio transients and also to establish a direct link to what we could expect for FRBs at higher frequencies, if they are indeed originated by magnetars. 

Although most pulsars display a steep spectral index, magnetars such as PSR 1745-2900, have a flat or inverted one \citep{torne2016,Pearlman_2018}. Their spectral index could also change in time (see e.g. \citealt{champion}). \cite{torne2015} reports a value of $\alpha=-0.4 \pm 0.1 $ for the GC magnetar, which changes to $\alpha=0.4\pm 0.2$ in a different epoch \citep{torne2016}. This positive value could reflect a turn up, related with the transition from coherent to incoherent radio emission. Observing with high-sensitivity instruments such as ALMA could be helpful to study the gap between centimetre and millimetre radio wavelengths, particularly with its new Band 1 set-up.

\subsection{PAM Band 1 predictions}

Based on our detections, we report a Poissonian rate of $\rm{R_{Band\hspace{1mm}3}}$ = 4 per hour (events/hr) for PSR J1745-2900 at ALMA Band 3. We can estimate the expected rate at Band 1 (42 GHz) using
\begin{equation}
    \rm{R_{Band\hspace{1mm}1} = R_{Band\hspace{1mm}3}\left(\frac{F_{Band\hspace{1mm}1}}{F_{Band\hspace{1mm}3}}\right)^{\gamma}},
\end{equation}
where $\rm{R_{Band-X}}$ corresponds to the expected event rate for a given fluence threshold at Band X. The fluence completeness limit using 37 antennas of 12m at Band 3 is $\rm{F_{Band\hspace{1mm}3}}$ = 1330 mJy ms and at Band 1 $\rm{F_{Band\hspace{1mm}1}}$ = 1200 mJy ms, considering a pulse width of 0.1 s. The exponent $\gamma$ is the power-law value derived from the cumulative energy distribution discussed in Sect. \ref{sec:energy_dist}.  We obtain a rate of  $\rm{R_{Band\hspace{1mm}1}} = 5 \pm 2$ events/hr. However, if we instead consider a total of 49 antennas for the phasing, then the expected rate rises to $\rm{R_{Band\hspace{1mm}1}} = 10 \pm 3$ events/hr, as the fluence threshold drops to $\rm{F_{Band\hspace{1mm}1}}$ = 900 mJy ms. Certainly, having a larger sample of pulses will provide more constraining values for their energy and frequency behaviour. In particular, measuring the spectral index of the source at different frequencies will tell us whether there is a turn-up point when we transition to incoherent emission mechanisms.

\subsubsection{FRBs}

If we consider the $\gamma = -2.4\pm0.1$ obtained for the GC magnetar and apply it to some actively repeating FRBs, we can estimate the event rates at ALMA bands. In this case, we re-calculated the fluence threshold for ALMA assuming bursts of 1 ms duration (typical for FRBs), and obtain $\rm{F_{Band-1}}$ = 90 mJy ms \footnote{Using a maximum of 49 antennas of the same 12m size.}. We take FRB\,20121102A, whose reported rates at L-band are 122 events/hr \citep{li2, li} and 218$\pm$16 events/hr \citep{NovRain}. This leads to a rate between 1 $\pm$ 1 and 3 $\pm$ 2 events/hr, respectively. If on the contrary, we take the flatter value of $\gamma=-1.1$ reported for FRB\,20121102A at L-band in \cite{papermary}, we estimate instead between 10 $\pm$ 3 and 32 $\pm$ 6 events/hr. It is noteworthy that these rates consider observations at the peak activity of the source.

We highlight that PAM Band 1 could serve as a bridge to millimetre wavelengths, which would have profound implications in constraining the FRBs' progenitor source. If detections are achieved, it will also be key to test the proposed frequency dependence of the activity window (chromaticity) of periodic FRBs. As discussed in \cite{surya}, the activity window seems to starts earlier and be narrower at higher frequencies. As shown in \href{fig:skyplot}{Fig. 4}, currently ALMA could potentially detect 19 repeating FRBs. This number is expected to increase rapidly given the new surveying instruments in the southern hemisphere such as MeerKAT \citep{meerkat}.

\subsubsection{Pulsars}

\begin{figure}
	\includegraphics[width=\columnwidth]{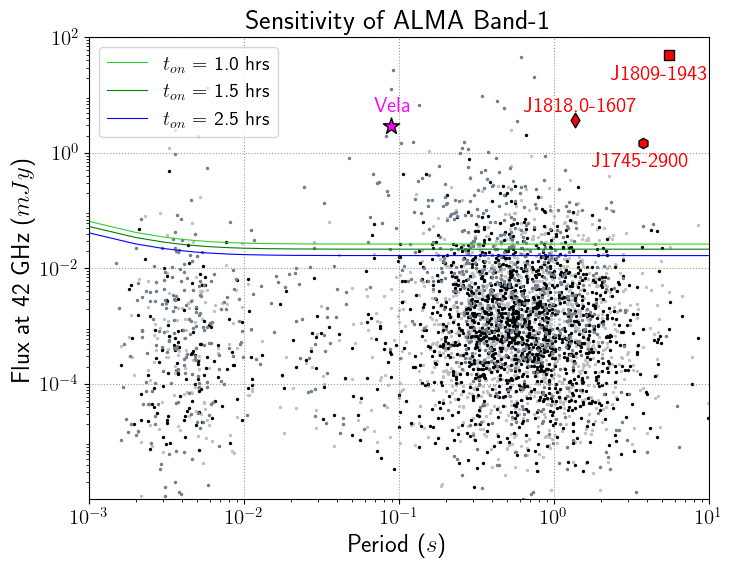}
    \caption{Predicted sensitivity of ALMA Band 1 to the known pulsar population. The black dots are the pulsars with reported spectral index, the grey dots are those without spectral index and from which we draw values from a normal distribution. The sources detectable lie above the threshold lines of integration times of 1 hr (light green), 1.5 hr (green) and 2.5 hr (blue). The light grey dots are pulsars not visible to ALMA due to the declination limit (< 47°). Based on Fig. 9 in \cite{Liu_2021}.} 
    \label{fig:band1_pulsars}
\end{figure}

We can also estimate the number of radio pulsars detectable in integrated pulse profile mode (also known as folding mode), which is obtained by adding hundreds to thousands of pulses using the spin period and other parameters listed in their ephemeris (see e.g. \href{tab:timing4}{Table 3}). We derived the minimum sensitivity for Band 3 using the radiometer equation for pulsars and considering telescope parameters such as $T_{sys}$ = 74 K, $G$ = 1.08 K/Jy \citep{almahand} for Band 3, a phasing efficiency of $\eta_{p} \sim 0.6$ \citep{matthews2017}, a minimum detected S/N of 6 and 1 hour of on-source integration time. For a pulsar spin period of 1 s with a duty cycle of 10\%, we estimate that the Band 3 system is sensitive to 0.060 mJy. If instead we use the parameters reported by \cite{Liu_2021}, we obtain 0.023 mJy. Assuming that the Band 1 system behaves similarly ($T_{sys}$ = 45 K, $G$ = 1.12 K/Jy in \citealt{almahand}), then we expect a minimum detectable flux of 0.035 mJy, using the same 37-antenna set-up. When increasing to 49 antennas, the minimum flux is 0.026 mJy. 

In \href{fig:band1_pulsars}{Fig. 5} we show the sources expected to be detectable by PAM Band 1. To this end, we selected all the known pulsars from the ATNF catalogue \citep{catalogue} having flux density reported at 1.4 GHz or 2 GHz. We limited the sources detectable by ALMA based on a declination below 47° \citep{almahand}. This reduced the list to 2301 pulsars.
 
We used the spectral index to estimate the flux of each source at 42 GHz (Band 1). To extrapolate we used a simple power-law:
 
\begin{equation}\label{eq:fluxeq}
    \rm{S_{42\hspace{0.6mm}GHz} = S_{1.4-2 \hspace{0.6mm}GHz}\left(\frac{\nu_{42\hspace{0.6mm}GHz}}{\nu_{1.4-2\hspace{0.6mm}GHz}}\right)^{\alpha}}.
\end{equation}
Here $\rm{S_{X}}$ corresponds to the flux at a given frequency in GHz, $\rm{\alpha}$ is the spectral index, and $\rm{\nu_{X}}$ is the frequency. For the sources that do not have a reported spectral index, we draw a spectral index from a normal distribution, considering a mean of -1.6 and standard deviation equal to 0.54 \citep{index}. We assumed a pulse width equal to 10$\%$ of the period and considered on-source integration times of 1, 1.5, and 2.5 hours. This would lead to the detection of 166 pulsars (7.2$\%$ of those visible to ALMA), 188 (8.2$\%$), and 209 (9.1$\%$), respectively. The sources detectable by PAM are the ones lying above the mentioned sensitivity lines in \href{fig:band1_pulsars}{Fig. 5}. Some remarkable detectable sources have been highlighted, such as the Vela pulsar, and three radio-loud magnetars, including PSR J1745$-$2900. It is important to mention that these surveys are unlikely to be sensitive to millisecond pulsars (MSPs, \citealt{Liu_2021}).

\section{Conclusions}\label{sec:sec6}

We reprocessed archival data of APS at Band 3 from Sgr A$^{*}$. We detected eight pulses from the GC magnetar PSR J1745$-$2900 and studied their cumulative energy distribution with the objective to link this with FRBs. The complex environment where the GC magnetar is located serves us to test promising progenitor scenarios for FRBs, given its high electron density and turbulent magneto ionic properties. 

We obtained an exponent of $\gamma$ = 2.4 $\pm$ 0.1 for the cumulative energy distribution of the GC magnetar. Using this exponent we predict a rate of 5 events/hr if we were to observe in Band 1, using 37 antennas of 12m. This rate increases to 10 events/hr with 49 12m antennas. For the periodic FRB~20121102A, at peak activity, we obtained a rate between 1 and 32 events/hr with PAM Band 1 using 49 12m antennas. Regarding pulsar detection, we estimated that there are roughly 166 pulsars potentially detectable within 1 hr integration. This number increases with more integration time.

 Currently, there are 19 repeating FRBs that are visible to ALMA. Given its high sensitivity and time resolution, instruments like PAM are crucial to exploring the FRB frequency extent to test their progenitor scenarios. Importantly, high-frequency searches are not affected by propagation effects, such as scattering, which could hinder a detection at cm-wavelengths. It is expected that the Band 1 receiver, including polarisation capabilities, will serve as a pioneer on mm-wavelengths searches for pulsars and to study complex environments involving radio transients. To cover the northern sky, telescopes such as the IRAM 30m or the NOEMA observatory could be used and, in the future, the ngVLA.

\begin{acknowledgements}
      A significant part of this work was developed during the NRAO REU Chile Internship, conducted at the Joint ALMA Observatory from January to March 2024. We extend our gratitude to AUI/NRAO Chile, D. Rebolledo and P. Cortés for their support. J.V-C gives special thanks to J.C. Fluxá for his support during the NRAO REU Chile Internship application, which ultimately led to this article. We also thank Prof. T. Cassanelli for his valuable feedback. This work made use of archival data from GMVA 2017 campaign with ALMA (Project Code: 2017.1.00795.V). We extend our gratitude to the P.I. M. Johnson and the team. This work is funded by the Max Planck Society through the Max Planck Partner Group led by Prof. M. Cruces. This work is funded by the Chinese Academy of Sciences President's International Fellowship Initiative. Grant No. 2026PVC0050).
\end{acknowledgements}

\bibliographystyle{aa}
\bibliography{aanda.bib}

\end{document}